\documentclass{article}
\usepackage{epsfig,amsmath,amssymb,enumitem}
\usepackage[papersize={8.5in,11in}]{geometry}
\begin{document}
\title{Identification of the 125 GeV Resonance as a Pseudoscalar Quarkonium Meson}
\author{J. W. Moffat\\~\\
Perimeter Institute for Theoretical Physics, Waterloo, Ontario N2L 2Y5, Canada\\
and\\
Department of Physics and Astronomy, University of Waterloo, Waterloo,\\
Ontario N2L 3G1, Canada}
\maketitle
\begin{abstract}
The 125 GeV resonance discovered at the LHC could be a heavy quarkonium pseudoscalar meson. The diagonalization of the mass matrix of the isoscalar quarkonium states $\vert\zeta\rangle$ and $\vert\zeta^{0'}\rangle$ produces the states identified with the pseudoscalar heavy quarkonium mesons $\zeta^0$ and $\zeta^{0'}$. For a mixing angle $\phi\sim 36\,^{\circ}$ the mass of the bound state pseudoscalar resonance $\zeta^0$ is $m_{\zeta^0}\sim 125$ GeV. The decay rates of the quarkonium $\zeta^0$ meson are estimated and compared to the standard model Higgs boson predictions. The importance of determining the spin-parity of the 125 GeV resonance is discussed. Criteria for experimentally discriminating between the pseudoscalar, $J^{PC}=0^{-+}$, $\zeta^0$ quarkonium meson and the scalar $J^{PC}=0^{++}$ Higgs boson are investigated.
\end{abstract}

\section{Introduction}

The search for the Higgs boson at the Large Hadron Collider (LHC) has discovered a new boson. The CMS~\cite{CMS} and ATLAS~\cite{ATLAS} experiments based on $5\,{\rm fb}^{-1}$ data show an excess of events at $\sim 125-126$ GeV that was evident already in the 2011 data. When combining the 7 TeV and 8 TeV data, both experiments separately reach a sensitivity corresponding to a local significance of $\sim 5\sigma$. The resonance could be the discovery of the elusive Higgs boson of the Standard Model (SM)~\cite{Djoudi}. However, we must make certain that the new resonance is the Higgs boson that was postulated to generate the electroweak symmetry breaking, which gives masses to the SM vector bosons $W$ and $Z$ and to the fermions. The LHC experiments will eventually have the sensitivity to accurately determine the decay rates for the final states $\gamma\gamma, ZZ^*, WW^*,Z\gamma,b\bar b, gg$ and $\tau^+\tau^-$ predicted by the Higgs boson production mechanism. The excess of events at $\sim 125$ GeV is mainly driven by the ``golden'' decay channels $\gamma\gamma$ and $ZZ^*\rightarrow 4$ leptons. In particular the latter decay channel is a significant signature of the Higgs coupling to a vector boson. However, the present sensitivity for this decay channel is about 3$\sigma$. There are anomalies in the decay channel observations, such as the zero or negative signal for $\tau^+\tau^-$ decay and that the signal size for the diphoton decay channel is about 1.5-2 times that predicted for the Higgs boson. Moreover, the data for the $b\bar b$ decay channel is consistent with the predicted Higgs boson rate.

In the following, we will attempt to identify the new boson with a neutral, heavy quarkonium bound state called $\zeta^0$~\cite{Moffat}.  The standard quantum chromodynamics (QCD) quark-gluon force binds the quarks and anti-quarks. An important test of the possible identification of the new boson as the $\zeta^0$ is its spin and parity, $J^{PC}=0^{-+}$. Due to the observed decay of the new boson into a diphoton state, it has to have either spin 0 or spin 2. Because the SM elementary Higgs boson is necessarily a scalar $J^{PC}=0^{++}$ boson with even parity, {\it it is critically important to determine the parity of the new boson}. This can be done at the LHC by analyzing the correlations of the angular distributions of the $\gamma\gamma$ and 4-lepton decay states of the 125 GeV boson.

\section{Mixed Quarkonium Resonance}

The bottomnium and unobserved toponium are isoscalar states $\vert B\rangle=\vert b{\bar b}\rangle$ and $\vert T\rangle=\vert t{\bar t}\rangle$ of heavy quarkonium. With respect to an effective interaction Hamiltonian, heavy quarkonium appears in two different isoscalar states $\vert\zeta^0\rangle$ and $\vert\zeta^{0'}\rangle$. The effective Hamiltonian is given by
\begin{equation}
{\cal H}_{\rm eff}={\cal H}_0+{\cal H}_{\rm mass},
\end{equation}
where
\begin{equation}
{\cal H}_{\rm mass}=K^T{\cal M}K.
\end{equation}
Here, ${\cal M}$ is the mass matrix:
\begin{equation}
\label{Massmatrix}
{\cal M}=\biggl(\begin{array}{cc}m_{\zeta'}&m_{\zeta\zeta'}\\
m_{\zeta\zeta'}&m_{\zeta}\end{array}\biggr),
\end{equation}
where $\vert\zeta\rangle$ and $\vert\zeta'\rangle$ are states of quarkonium that interact through the mixing contributions $m_{\zeta\zeta'}$ and $K=\biggl(\begin{array}{cc}\zeta'\\\zeta\\\end{array}\biggr)$.
The mass matrix can be diagonalized:
\begin{equation}
\label{Diagonalmatrix}
D=R^T{\cal M}R,
\end{equation}
where $R$ is the rotation matrix:
\begin{equation}
\label{rotationmatrix}
R=\biggl(\begin{array}{cc}\cos\phi&\sin\phi\\-\sin\phi&\cos\phi\end{array}\biggr),
\end{equation}
and $D$ is the diagonal matrix:
\begin{equation}
D=\biggl(\begin{array}{cc}m_T&0\\0&m_B\end{array}\biggr).
\end{equation}
Here, $m_T\sim 2m_t\sim 346$ GeV and $m_B\sim 2m_b\sim 9$ GeV where we have used the measured quark masses: $m_t=173$ GeV and $m_b\sim 4.5$ GeV.

In terms of the states $\vert T\rangle$ and $\vert B\rangle$, we have
\begin{equation}
{\cal H}'_{\rm mass}=m_TTT^\dagger+m_BBB^\dagger,
\end{equation}
where
\begin{equation}
\biggl(\begin{array}{cc}T\\B\\\end{array}\biggr)
=R\biggl(\begin{array}{cc}\zeta'\\\zeta\\\end{array}\biggr)
=\biggl(\begin{array}{cc}\cos\phi \zeta'+\sin\phi \zeta\\\cos\phi\zeta-\sin\phi\zeta'\end{array}\biggr).
\end{equation}
From (\ref{Diagonalmatrix}), we obtain the masses:
\begin{equation}
m_T=\cos^2\phi m_{\zeta'}-2\sin\phi\cos\phi m_{\zeta\zeta'}+\sin^2 m_\zeta,
\end{equation}
and
\begin{equation}
m_B=\cos^2\phi m_\zeta+2\sin\phi\cos\phi m_{\zeta\zeta'}+\sin^2 m_{\zeta'}.
\end{equation}

By inverting (\ref{Diagonalmatrix}), we get
\begin{equation}
{\cal M}=RDR^T
\end{equation}
which leads to the result
\begin{equation}
\label{zetasolution}
{\cal M}\equiv\biggl(\begin{array}{cc}m_{\zeta'}&m_{\zeta\zeta'}\\
m_{\zeta\zeta'}&m_\zeta\end{array}\biggr)=\biggl(\begin{array}{cc}\cos^2\phi m_T+\sin^2\phi m_B&\cos\phi\sin\phi(m_B-m_T)\\
\cos\phi\sin\phi(m_B-m_T)&\cos^2\phi m_B+\sin^2\phi m_T\end{array}\biggr).
\end{equation}
We derive from (\ref{zetasolution}) the results
\begin{equation}
\label{zetamass}
m_\zeta=\cos^2\phi m_B+\sin^2\phi m_T,
\end{equation}
and
\begin{equation}
\label{zetaprimemass}
m_{\zeta'}=\cos^2\phi m_T+\sin^2\phi m_B.
\end{equation}
The off-diagonal term is given by
\begin{equation}
m_{\zeta\zeta'}=\cos\phi\sin\phi(m_B-m_T).
\end{equation}

We can determine the mixing angle $\phi$ from the equation:
\begin{equation}
\label{mixingangle}
\phi=\arccos[(m_T-m_{\zeta})/(m_T-m_B)]^{1/2}.
\end{equation}
The mixing angle $\phi\sim 36\,^{\circ}$ is obtained from (\ref{mixingangle}) and from (\ref{zetamass}) and (\ref{zetaprimemass}), we get the masses of the quarkonium states $\vert\zeta\rangle$ and $\vert\zeta'\rangle$: $m_{\zeta^0}\sim 125$ GeV and $m_{\zeta^{0'}}\sim 230$ GeV. We identify the new boson resonance discovered at the LHC with the $\zeta^0$ bound state quarkonium resonance. The quarks and anti-quarks are bound together by the QCD gluon force with a corresponding binding energy $E_B$. A possible non-relativistic heavy quark gluon potential is
\begin{equation}
V=-\frac{a}{r}+br,
\end{equation}
and the QCD coupling constant $\alpha_s(M_Z)=0.118$~\cite{pdg}.

The short life-time of the top quark for the decay $t\rightarrow bW^+$, $\tau_t\sim 5\times 10^{-25}$ sec., results in the toponium not forming a bound state. However, the toponium state exists as an state of heavy quarkonium. The heavy $\zeta^{0'}$ with a mass $m_{\zeta^{0'}}\sim 328$ GeV does not form a detectable bound state.

\section{Prominent Decays of $\zeta^0$ and Higgs Boson}

The $\zeta^0$ meson will be primarily produced at the LHC in gluon-gluon fusion with the largest cross section for the $^1{S_0}$ ground state, and this state will decay into $gg,\gamma\gamma, c\bar c, b\bar b$, $\tau^+\tau^-,\mu^+\mu^-,WW^*$ and $ZZ^*$. At leading order and in the narrow-width approximation, the production cross section for a boson $B$ in $pp$ collisions is given in terms of the gluon decay width by
\begin{equation}
\label{prodcrosssection}
\sigma(pp\rightarrow B+X)=\frac{\pi^2}{8m_B^3}\Gamma(B\rightarrow gg)\int^1_{\tau_B} dx\frac{\tau_B}{x}g(x,Q^2)g(\tau_B/x,Q^2),
\end{equation}
where $g(x,Q^2)$ is the gluon parton distribution function, and $\tau_B=m_{B}^2/s$ where $s$ is the pp collision energy squared. We identify $B$ with either the pseudoscalar meson $B=\zeta^0$ or the Higgs particle $B=H^0$.

We can obtain approximate results for the $\zeta^0$ boson decay widths, using the colored non-relativistic heavy quark model~\cite{REllis}. The quark wave function for the lowest lying S-wave state is
\begin{equation}
\psi(r)=\frac{1}{\sqrt{4\pi}}R_0(r).
\end{equation}
In the Coulomb approximation, we have
\begin{equation}
\frac{\vert R_0(0)\vert^2}{m_{\zeta^0}}=C_i{\alpha_s}^3m_{\zeta^0},
\end{equation}
where $C_i$ is the quark charge and color factor. The parametrization of the bound state wave functions, their binding energy and the $R(0)$ at the origin have been computed in ref.~\cite{Hagiwara}.

The form for the two-particle $\zeta^0$ annihilation decay width is~\cite{Drees,Martin}
\begin{equation}
\Gamma(\zeta^0\rightarrow AB)
=\frac{3}{32\pi^2(1+\delta_{AB})}N^{1/2}(1,m_A^2/m_{\zeta}^2,m_B^2/m_{\zeta}^2)\frac{\vert R_0(0)\vert^2}{m_{\zeta}^2}\sum\vert M\vert^2.
\end{equation}
Here, $M$ are the amplitudes and $AB=gg,\gamma\gamma,ZZ^*,WW^*,Z\gamma,b{\bar b},c{\bar c}$ and $\delta_{AB}=1$ for the cases $gg,\gamma\gamma,ZZ^*$. Moreover, $N(x,y,z)=x^2+y^2+z^2-2xy-2xz-2yz$. The production cross section in proton-proton collisions for $\zeta^0\rightarrow\gamma\gamma$ is given by
\begin{equation}
\sigma(pp\rightarrow\zeta^0\rightarrow\gamma\gamma)=\frac{\pi^2}{8m_{\zeta}^3}\Gamma(\zeta^0\rightarrow\gamma\gamma)
\int_\tau^1dx\frac{\tau}{x}g(x,Q^2)g(\tau/x,Q^2),
\end{equation}
where $BR(\zeta^0\rightarrow\gamma\gamma)$ is the branching ratio. The ratio of the $\zeta^0\rightarrow\gamma\gamma$ production cross section to that of the SM model Higgs boson $H^0$ is
\begin{equation}
\mu=\frac{\sigma(\zeta^0\rightarrow\gamma\gamma)}{\sigma(H^0\rightarrow\gamma\gamma)}.
\end{equation}
For the mass $m_\zeta=m_H\sim 125$ GeV the ratio $\mu$ could be greater than unity in agreement with the current excess signal for the diphoton decay of the new boson~\cite{CMS,ATLAS}.

The $M$ for the $W^+W^-$ final state is given in the heavy quarkonium zero velocity approximation $v\sim 0$ by
\begin{equation}
\label{WWamplitude}
M^{\lambda{\bar\lambda}}(\zeta^0\rightarrow W^+W^-)=-\gamma_W^{2-\vert\lambda\vert-\vert{\bar\lambda}\vert}
(\delta_{\lambda 0}\delta_{{\bar\lambda}0}+(-)^{\lambda}\delta_{\lambda{\bar\lambda}})
\frac{1}{2}g^2\cos^2\theta-2\beta_W^2\gamma_W^2\delta_{\lambda 0}\delta_{{\bar\lambda 0}}g^2\cos^2\theta\frac{m_\zeta^2}{m_\zeta^2-m_W^2}.
\end{equation}
where $\lambda,{\bar\lambda}$ are specific helicities of the $W$ bosons, which can take on the values $0,\pm 1$. Moreover, $\theta$ is the t-channel scattering angle, $g$ and $\sin\theta_W$ are the SM gauge parameters, $\gamma_W=m_\zeta/m_W$ and $\beta_Z=\sqrt{1-(m_W/m_\zeta)^2}$.

The $M$ for the $ZZ$ final state is given in the zero velocity approximation $v\sim 0$ by
\begin{equation}
\label{ZZamplitude}
M^{\lambda{\bar\lambda}}(\zeta^0\rightarrow ZZ)=-\gamma_Z^{2-\vert\lambda\vert-\vert{\bar\lambda}\vert}
(\delta_{\lambda 0}\delta_{{\bar\lambda}0}+(-)^{\lambda}\delta_{\lambda{\bar\lambda}})
\frac{1}{\cos\theta_W}\biggl[2g^2\biggl(\biggl(\frac{1}{4}-\frac{2}{3}\sin^2\theta_W\biggr)\cos^2\theta
+\frac{4}{9}\sin^4\theta_W\biggr)\biggr]
$$ $$
+\frac{2g^2m_\zeta^2}{\cos^2\theta_W}\beta_Z^2\gamma_Z^2\delta_{\lambda 0}\delta_{{\bar\lambda}0}\biggl[\frac{(\cos^2\theta-\frac{4}{3}\sin^2\theta_W)^2}{2m_\zeta^2-m_Z^2}
+\frac{\cos^2\theta\sin^2\theta}{m_\zeta^2-m_Z^2}\biggr],
\end{equation}
where $\gamma_Z=m_\zeta/m_Z$ and $\beta_Z=\sqrt{1-(m_Z/m_\zeta)^2}$.

Finally, for the $Z\gamma$ final state we get
\begin{equation}
\vert M(\zeta^0\rightarrow Z\gamma)\vert^2=\frac{32\pi Q_q^2\alpha g^2}{\cos^2\theta_W}\biggl(\frac{1}{2}\cos^2\theta-\frac{2}{3}\sin^2\theta_W\biggr)^2,
\end{equation}
where $Q_q$ is the quark charge and only $\lambda_Z=\lambda_\gamma=\pm 1$ helicity values are permitted.

As in the SM with a Higgs boson, the coupling of the $\zeta^0$ to $Z$ and $W$ bosons is stronger than the coupling to photons, and we have that $\Gamma(\zeta^0\rightarrow ZZ^*) > \Gamma(H^0\rightarrow\gamma\gamma)$.

For the S-wave initial $\zeta^0$ state and $v\rightarrow 0$, we have for the squared amplitudes for the $gg$ and $\gamma\gamma$ final states:
\begin{equation}
\label{ggamplitude}
\sum_{{\rm spins},{\rm color}}\vert M(\zeta^0\rightarrow gg)\vert^2=S_{gg}Q_q(4\pi\alpha_s)^2
\end{equation}
and
\begin{equation}
\label{photonamplitude}
\sum_{{\rm spins}}\vert M(\zeta^0\rightarrow\gamma\gamma)\vert^2=S_{\gamma\gamma}(4\pi\alpha)^2,
\end{equation}
where $S_i$ are spin and color factors. We find from the ratio of (\ref{ggamplitude}) and (\ref{photonamplitude}) that $\Gamma(\zeta^0\rightarrow gg)\gg\Gamma(\zeta^0\rightarrow\gamma\gamma)$.

For the decay $\zeta^0\rightarrow \gamma\gamma$, we get to leading order in the Born approximation:
\begin{equation}
\label{pseudogammadecay}
\Gamma(\zeta^0\rightarrow \gamma\gamma)=C_{\gamma\gamma}\biggl(\frac{\alpha}{m_{\zeta^0}}\biggr)^2\vert R_0(0)\vert^2= C_{\gamma\gamma}\alpha^2{\alpha_s}^3m_{\zeta^0}.
\end{equation}

We can compare the leading order partial decay width of $\zeta^0\rightarrow \gamma\gamma$ to the partial width of the scalar Higgs particle decay in the light mass Higgs limit~\cite{JEllis,Shifman,Marciano}:
\begin{equation}
\Gamma(H^0\rightarrow\gamma\gamma)=\vert I\vert^2\biggl(\frac{\alpha}{4\pi}\biggr)^2\frac{G_Fm_H^3}{8\sqrt{2}\pi},
\end{equation}
where $G_F=1.6637\times 10^{-5}\,{\rm GeV^{-2}}$ is Fermi's constant. In the limit of a Higgs mass $m_H\sim 125$ GeV and for the W and top quark loop contributions, $I_{(W+t)}\sim - 10$, we obtain
\begin{equation}
\label{Higgsgamma}
\Gamma(H^0\rightarrow \gamma\gamma)\sim 31\, {\rm keV}.
\end{equation}

For the gluon final state, we get for the $\zeta^0$ meson decay:
\begin{equation}
\label{gluons}
\Gamma(\zeta^0\rightarrow\, gg)=\frac{4}{3}C_{gg}\biggl(\frac{\alpha_s}{m_{\zeta^0}}\biggr)^2\vert R_0(0)\vert^2= \frac{4}{3}C_{gg}{\alpha_s}^5m_{\zeta^0}.
\end{equation}
The partial decay width in the Born approximation for the decay $H^0\rightarrow gg$ is given
by
\begin{equation}
\label{Higgsgluons}
\Gamma(H^0\rightarrow gg)=\frac{G_F{\alpha}_s^2m_H^3}{36\sqrt{2}\pi^3}.
\end{equation}

Measurements of the total production cross sections $\sigma_\zeta$ and $\sigma_H$ and the branching ratios of decay channels at the LHC can be used to distinguish between the $\zeta^0$ pseudoscalar heavy quarkonium resonance and the SM Higgs boson.

\section{Parity in the Diphoton Decay of $\zeta^0$}

Let us consider the non-relativistic quarkonium state as the product of the orbital wave function and the spin vector $S$~\cite{Harpen}:
\begin{equation}
\Psi_{n.l,m}({\bf r}\vert S,S_z\rangle.
\end{equation}
The spin vectors are linear combinations of products of the individual quarks:
\begin{equation}
\vert S=1,S_z=1\rangle=\vert \uparrow\rangle\vert\uparrow\rangle,
$$ $$
\vert S=1, S_z=0\rangle=\frac{1}{\sqrt{2}}(\vert\uparrow\rangle\vert\downarrow\rangle+\vert\downarrow
\rangle\vert\uparrow\rangle),
$$ $$
\vert S=1, S_z=-1\rangle=\vert\downarrow\rangle\vert\downarrow\rangle,
$$ $$
\vert S=0, S_Z=0\rangle=\frac{1}{\sqrt{2}}(\vert\uparrow\rangle\vert\downarrow\rangle
-\vert\downarrow\rangle\vert\uparrow\rangle),
\end{equation}
where $\vert\uparrow\rangle\vert\uparrow\rangle$ is the product of an anti-quark spin state with $S_z=+1/2$ and a quark state with $S_z=+1/2$, producing an state of total spin $S=1$ and $z$ component $S_z=1$. In analogy with the states of positronium, the first three expressions above describe the possible three states of ``ortho-quarkonium'', while the last expression describes the spin state of ``para-quarkonium''.

By applying the parity operator to the quarkonium orbital wave function, we get
\begin{equation}
P\Psi_{n,l,m}({\bf r})=\Psi_{n,l,m}(-{\bf r})=(-1)^l\Psi_{n,l,m}(\bf r).
\end{equation}
The intrinsic parity of a quark is the opposite of the anti-quark. The overall parity of a quarkonium state is $-(-1)^l$ and since both ortho- and para-quarkonium decay from a $l=0$ state both have odd $-1$ parity at decay. Since parity is conserved in electromagnetic interactions, the photon states must have odd parity. The wave function of the photon with momentum ${\bf k}$ and transverse polarization $\epsilon$ is
\begin{equation}
A({\bf r},t)=\exp(i({\bf k}\cdot{\bf r}-\omega t)\vert\epsilon\rangle.
\end{equation}
The symmetrized wave function for the two-photon state with momenta ${\bf k}$ and $-{\bf k}$ and polarizations $\bf\epsilon_1$ and $\bf\epsilon_2$ is given by
\begin{equation}
A({\bf r},t)=\exp(i({\bf k}\cdot({\bf r}-2\omega t)\vert\epsilon_1\rangle\vert\epsilon_2\rangle+\exp(i(-{\bf k}\cdot({\bf r}-2\omega t)\vert\epsilon_2\rangle\vert\epsilon_1\rangle,
\end{equation}
where ${\bf r}={\bf r}_1-{\bf r}_2$ and $\vert\epsilon_1\rangle$ and $\vert\epsilon_2\rangle$ are orthogonal. This expression can be rewritten as
\begin{equation}
\label{photonparity}
A({\bf r},t)
=\frac{1}{2}[\vert\epsilon_1\rangle\vert\epsilon_2\rangle+\vert\epsilon_2\rangle\vert\epsilon_1\rangle]
\{\exp(i({\bf k}\cdot{\bf r}-2\omega t)+\exp(i(-{\bf k}\cdot{\bf r}-2\omega t)\}
$$ $$
+\frac{1}{2}[\vert\epsilon_1\rangle\vert\epsilon_2\rangle-\vert\epsilon_2\rangle\vert\epsilon_1\rangle]
\{\exp(i({\bf k}\cdot{\bf r}-2\omega t)-\exp(i(-{\bf k}\cdot{\bf r}-2\omega t)\}.
\end{equation}
In the last equation the terms both exhibit Bose symmetry of the interchange of the Bose indices, $1\longleftrightarrow 2$, and the first term has even and the second term odd parity. The para-quarkonium decay state is described by the second term in (\ref{photonparity}):
\begin{equation}
A_{\rm PQ}({\bf r},t)
=\frac{1}{\sqrt{2}}[\vert\epsilon_1\rangle\vert\epsilon_2\rangle-\vert\epsilon_2\rangle\vert\epsilon_1\rangle]
\{\exp(i({\bf k}\cdot{\bf r}-2\omega t)-\exp(i(-{\bf k}\cdot{\bf r}-2\omega t)\}.
\end{equation}
This polarization state is a superposition of different polarization states of the individual photons. Nonetheless, the final measured value of the polarization state is fixed by conservation of parity and the orthogonality of the polarization vectors.

\section{Parity of Decays to ZZ}

The scalar character of the Higgs boson with $J^{PC}=0^{++}$ can be tested at the LHC. Below the threshold for two real $Z$ bosons, the Higgs particle can decay into real and virtual $WW^*$ and $ZZ^*$ pairs. The partial decay width for the Higgs decay into $ZZ$ is given by~\cite{Rizzo}:
\begin{equation}
\Gamma(H^0\rightarrow ZZ)=\frac{3G_Fm_H^3}{32\pi}(1-4x+12x^2)\sqrt{(1-4x)},
\end{equation}
where $x=m_Z^2/m_H^2$. The parity of the SM Higgs particle is necessarily $P=+1$. However, a pseudoscalar coupling to a Higgs-type particle can be achieved in two-doublet models, in which the coupling to $ZZ$ is induced by higher-order loop effects~\cite{Gunion}. Moreover, supersymmetry models have multiple Higgs bosons of which one can be a pseudoscalar boson $A$.

The way to test the parity and spin of the 125 GeV boson is to determine the angular correlations in the boson decay to fermion and gauge boson pairs. The observation by the CMS and ATLAS groups of the decay or fusion process into two photons rules out the $J=1$ directly by the Landau-Yang theorem~\cite{Landau,Yang} and determines that $C=+1$. This would also follow from the decay of the boson into $WW^*$ and $ZZ^*$. The longitudinal wave function of a vector boson grows with energy in contrast to the transverse wave function. The $Z$ boson in the S-wave Higgs production process becomes asymptotically longitudinally polarized. By contrast, the $Z$ boson in the $\zeta^0$ production process becomes transversally polarized at high energies. This experimental signature can be checked at the LHC. The distribution of lepton pairs in the $Z\rightarrow \ell^-\ell^+$ rest frame relative to the flight direction of the $Z$ is given by $\sin^2\theta$ for longitudinally polarized $Z$ bosons, while for transversally polarized states, after averaging over azimuthal angles, it behaves as $(1\pm\cos^2\theta)^2$.

The azimuthal angular correlations as well as the polar angles depend sensitively on the spin-parity assignments of the boson particle. We will study the scalar $J^{PC}=0^{++}$ Higgs and pseudoscalar $J^{PC}=0^{-+}$ $\zeta$ boson decays~\cite{Barger,Choi,Gao,Rujula}:
\begin{equation}
B\rightarrow ZZ^*\rightarrow (\ell_1^-\ell_1^+)(\ell_2^-\ell_2^+),
\end{equation}
where $B=[H^0,\zeta^0]$. The reconstructed $Z$ bosons are back-to-back in the $B$ boson rest frame. The two $Zs$ are labeled $Z_1$ and $Z_2$. The reconstructed masses of the $Z$ bosons are denoted by $m_1$ and $m_2$, and they define the longitudinal boosts from the $B$ rest frame to the rest frames of the decaying $Z_1$ and $Z_2$ bosons:
\begin{equation}
\gamma_1=\frac{m_B}{2m_1}\biggl(1+\frac{m_1^2-m_2^2}{m_B^2}\biggr)
\end{equation}
and
\begin{equation}
\gamma_2=\frac{m_B}{2m_2}\biggl(1-\frac{m_1^2-m_2^2}{m_B^2}\biggr).
\end{equation}
We denote by $\theta_1,\phi_1$ and $\theta_2,\phi_2$ the $\ell_1^-$ and $\ell_2^-$ decay angles in the $Z_1$ and $Z_2$ rest frames, respectively.

The decay amplitudes are defined in terms of the two boost parameters $\gamma_1$ and $\gamma_2$:
\begin{equation}
\gamma_a=\gamma_1\gamma_2(1+\beta_1\beta_2)
\end{equation}
\begin{equation}
\gamma_b=\gamma_1\gamma_2(\beta_1+\beta_2),
\end{equation}
where the $\beta_i,\gamma_i$ are the velocities and $\gamma$ factors of the on/off-shell $Z$ bosons. The $\gamma_a$ and $\gamma_b$ satisfy the condition
\begin{equation}
\gamma_a^2-\gamma_b^2=1.
\end{equation}
We have
\begin{equation}
\gamma_a=\frac{1}{2m_1m_2}[m_B^2-(m_1^2+m_2^2)].
\end{equation}

Effective Lagrangians for the $HZZ$ and $\zeta ZZ$ couplings are
\begin{equation}
{\cal L}_{HZZ}=(\sqrt{2}G_F)^{1/2}m_Z^2HZ^\mu Z_\mu,
\end{equation}
and
\begin{equation}
\label{Pwave}
{\cal L}_{\zeta ZZ}=hm_Z^2\zeta Z^{\mu\nu}{\tilde Z}_{\mu\nu}.
\end{equation}
Here, $Z_{\mu\nu}=\partial_\mu Z_\nu-\partial_\nu Z_\mu$, $h$ is a coupling constant, ${\tilde Z}^{\mu\nu}=\epsilon^{\mu\nu\rho\sigma}Z_{\rho\sigma}$, and (\ref{Pwave}) is a P-wave coupling, odd under parity and even under charge conjugation.

The angular distribution of the leptons in $H^0\rightarrow ZZ\rightarrow 4l$ decay, for on or off-shell $Zs$ of mass $m_1$ and $m_2$ for the scalar Higgs boson is given by~\cite{Rujula}:
\begin{equation}
\frac{d\Gamma(H^0\rightarrow ZZ^*)}{d\cos\theta_1d\cos\theta_2d\phi}\propto m_1^2m_2^2m_H^4[1+\cos^2\theta_1\cos^2\theta_2+(\gamma_b^2+\cos^2\phi)\sin^2\theta_1\sin^2\theta_2
$$ $$
$$ $$
+2\gamma_a\cos\phi\sin\theta_1\sin\theta_2\cos\theta_1\cos\theta_2+2\eta^2(\cos\theta_1\cos\theta_2
+\gamma_a\cos\phi\sin\theta_1\sin\theta_2)].
\end{equation}
Here, $\eta$ denotes the quantity occurring in the SM couplings of the $Z$ bosons to the final state leptons:
\begin{equation}
\eta=\frac{2c_vv_a}{(c_v^2+c_a^2)}\sim 0.15.
\end{equation}

For the pseudoscalar $\zeta^0$ decay $\zeta^0\rightarrow ZZ^*\rightarrow (\ell_1^-\ell_1^+)(\ell_2^-\ell_2^+)$, we obtain~\cite{Rujula}:
 \begin{equation}
\frac{d\Gamma(\zeta^0\rightarrow ZZ^*)}{d\cos\theta_1d\cos\theta_2d\phi}\propto m_1^4m_2^4\gamma_b^2(1+\cos^2\theta_1\cos^2\theta_2-\cos^2\phi\sin^2\theta_1\sin^2\theta_2+2\eta^2\cos\theta_1\cos\theta_2).
\end{equation}

The general feature that distinguishes the scalar Higgs boson decay from the pseudoscalar $\zeta^0$ decay is that for scalar particles the production amplitude $\sim{\vec\epsilon}_1\cdot{\vec\epsilon}_2$ is non-zero for parallel polarization vectors, while for the pseudoscalar $\zeta^0$ the amplitude $\sim{\vec\epsilon}_1\times {\vec\epsilon}_2$ is non-zero for perpendicular polarization vectors. The zero spin of the new boson at 125 GeV can be checked experimentally by observing the lack of any correlation between the final and initial state particles.

\section{Conclusions}

We have developed a model in which the newly discovered boson at the LHC with a mass $125-126$ GeV can be identified with a heavy quarkonium pseudoscalar resonance $\zeta^0$. By mixing the heavy quarkonium states $\vert\zeta^0\rangle$ and $\vert\zeta^{0'}\rangle$ through a rotation angle $\phi\sim 36\,^{\circ}$, we obtain two states $\vert\zeta^0\rangle$ and $\vert\zeta^{0'}\rangle$ with the masses $m_{\zeta^0}\sim 125$ GeV and $m_{\zeta^{0'}}\sim 230$ GeV. The $\zeta^0$ can form a bound state with the standard QCD gluon interaction. The rates of decays into light hadrons and leptons predicted by the $\zeta^0$ can be compared to the decay rates of the Higgs boson, so that it can be verified which boson has been observed at the LHC. Moreover, the important signature of the differences in the parity of the $\zeta^0$ and Higgs boson can eventually be determined at the LHC, when enough data and a large enough luminosity have been obtained. This will unambiguously distinguish between two of the possible identifications of the new boson as a heavy quarkonium meson or a SM Higgs boson.

\section*{Acknowledgements}

I thank Viktor Toth, Alavaro De Rujula, Bob Holdom, Philip Mannheim and Martin Green for helpful discussions. This research was generously supported by the John Templeton Foundation. This research was supported in part by Perimeter Institute for Theoretical Physics. Research at Perimeter Institute is supported by the Government of Canada through Industry Canada and by the Province of Ontario through the Ministry of Economic Development and Innovation.

\end{document}